\documentclass[aps,floatfix,twocolumn]{revtex4} 
\usepackage{graphicx,tikz}
\usepackage{amsmath}
\def\be{\begin{equation}}
\def\ee{\end{equation}}
\def\bea{\begin{eqnarray}}
\def\eea{\end{eqnarray}}
\def\bd{\begin{displaymath}}
\def\ed{\end{displaymath}}

\def\ga{\gamma}
\def\etal{{\em et al. }}

\begin{document}
\title{Neutron Capture Reactions near $N=82$ Shell-closure}
\author{Saumi Dutta}
\email{saumidutta89@gmail.com}
\author{Dipti Chakraborty}
\email{diptichakraborty2009@gmail.com}
\author{G. Gangopadhyay}
\email{ggphy@caluniv.ac.in}
\author{Abhijit Bhattacharyya}   
\email{abhattacharyyacu@gmail.com} 
\affiliation{Department of Physics, University of Calcutta\\
92, Acharya Prafulla Chandra Road, Kolkata-700 009, India}
\textwidth 7in
\textheight 8in

\begin{abstract}
Neutron capture cross-sections have been calculated in nuclei near the 
$N=82$ neutron shell closure. These nuclei are of astrophysical interest, participating
in s-process and p-process. A semi-microscopic optical model have been used
with the potential being obtained through folding the target density with the 
DDM3Y nucleon nucleon interaction.  Theoretical density values have been calculated
in the relativistic mean field approach. The calculated cross-section, as a function of 
neutron energy, agree reasonably well with experimental measurements. Maxwellian averaged 
cross-sections, important for astrophysical processes, have been calculated.
\end{abstract}
\maketitle

\section{Introduction}

After the seminal work of Burbidge \etal \cite{B2FH}, the method of creation of 
heavy elements through the slow neutron capture or s-process has been 
firmly established. A recent review that discusses our current understanding of the 
s-process may be found in K\"{a}ppeler \etal \cite{review} It is now understood that 
a major fraction of the nuclear abundance in the mass region
$90<A<209$ is due to the main s-process which takes place in He shells in 
low-mass AGB stars of moderate mass.

The role of the neutron capture reaction in s-process has been explored in 
many works. More accurate measurements have shown the inadequacy of the classical 
site-independent s-process and have paved the way to coupling with stellar models.
A mass region, where the reaction cross-section plays a very important role,
lies near the $N=82$ shell closure. Here, the elements Ba ($Z=56$) to Sm ($
Z=62)$ have small cross-sections 
for neutron capture reactions because of the proximity of the  shell-closure.
Hence, they act as bottle-necks for the s-process reaction path.   
An s-process peak occurs at  $^{138}$Ba. There are several s-only nuclides
such as $^{134,136}$Ba, $^{142}$Nd, and $^{148,150}$Sm in this mass region.
In the case of pairs of s-only isotopes such as $^{134,136}$Ba, the cross-sections and 
the abundances can be used to obtain the branching ratios of the s-process. Nuclei on the s-process
path that have comparable beta decay rates and neutron capture rates 
act as branch points as the nucleosynthesis path bifurcates towards both the proton
and neutron rich sides while passing through them.
The cross-sections at the branch points and s-only isotopes can provide 
important clues to the physical environments where the s-process takes place.

Although  experimental measurements are available for many isotopes in the 
mass region, cross-section values are required for some unstable nuclei  
important to determine the branching ratios. Such nuclei include
$^{134,135}$Cs, $^{141}$Ce, $^{147}$Nd, and $^{147,148}$Pm. 
We should also remember that although the classical or canonical s-process
calculations use the Maxwellian averaged cross-sections (MACS) at a single 
thermal energy ($\approx$ 30 keV usually), recent approaches, which couple stellar
models with the s-process network, need MACS at different thermal energies. 
Measurements are not always available and extrapolation to too distant 
value from the measured ones may lead to errors. Theoretical calculations can 
supplement the experimental measurements in this regard.

There are some neutron capture reactions in this mass region whose studies are relevant for the
astrophysical p-process. 
Photodissociation reactions such as $(\ga,n)$ occur in extremely hot environments.
Explosively burning Ne/O layer in core-collapse supernovae
heated by the outgoing shock-wave may provide such an environment.
Cross section values are very important as various photo-dissociation reactions such as 
$(\ga,n)$, $(\ga,p)$ and $(\ga,\alpha)$ compete at high temperature.
The emitted neutrons also may be absorbed after the shock wave passes through the 
layer. 
Thus it is very important to measure
the cross-sections for relevant  $(\ga,n)$ reactions at thermal energies. The reverse process, i.e., $(n,\gamma)$ reactions may serve the purpose.
For example, Dillmann \etal \cite{p-proc} studied a number of $(n,\ga)$ reactions with neutrons from
the $^7$Li($p,n$)$^7$Be reaction to simulate a Maxwellian neutron distribution 
at 25 keV thermal energy.

Neutron capture reactions have been studied in various methods. Older experiments usually used 
neutron beams of comparatively wide resolution. As the resonances in this region are narrow
(less than 1 eV), one gets an average cross-section in such experiments. Extremely high resolution
experiments using the neutron time of flight technique (TOF) have been used to study the
resonances. However, we are more interested in the Maxwellian averaged cross-sections.
In such cases, the data have been compressed into coarse energy bins to obtain
MACS values. In some other
experiments, sources of neutrons have been used that closely simulated the thermal neutron 
spectrum at certain temperatures. They can provide direct measurement of MACS values.

In the present work, we have studied low energy neutron capture cross-sections
of the nuclei near $N=82$. In the next section, we briefly present our 
formalism. In Section III, we discuss our results for important neutron capture 
reactions near $N=82$ shell closure, first at various neutron energies and 
compare with experimental measurements. It is followed by a calculation of
the MACS values, first at 30 keV, and later at different thermal energies for 
some selected isotopes. Finally we summarize our work.

\section{Calculation}

We have used a semi-microscopic procedure to calculate the neutron capture cross-section in the
present work. This method has been followed in a number of our recent works 
 \cite{GG, Chirashree1, Chirashree2, Chirashree3, Saumi, Dipti}. For example, in 
Chakraborty \etal \cite{Dipti}, this procedure has been utilized to study proton capture reactions, 
important for the astrophysical p-process in mass $110-125$ region. In the present approach, we 
extend it to study neutron capture reactions near $N=82$ shell closure. 

To briefly describe our procedure, we have 
assumed spherical symmetry for the target nuclei. The density profiles of the nuclei have been 
calculated in co-ordinate space in the relativistic mean field (RMF) approach using the 
FSU Gold Lagrangian density \cite{fsugold}.  
The charge density is obtained from the point proton density considering  the
finite size of the proton using a standard Gaussian form factor $F(r)$ \cite{form}
as follows. 
\begin{equation}
\rho(r)=e\int \rho ({\bf r\prime})F({\bf r}-{\bf r\prime}) d {\bf r\prime}
\end{equation}
\begin{equation}
F(r)=(a \sqrt\pi)^{-3}exp(-r^{2}/a^{2})
\end{equation}
with $a=\sqrt{2/3}a_p$, where $a_p=0.80$ fm is the root mean square (r.m.s.)
charge radius of  the proton. 
The charge density thus obtained is used to calculate r.m.s
charge radii for some nuclei in and around the concerned region of shell closure
to compare with experimentally available values. Comparison with measured
values serves as a check on
the applicability and reliability of the Lagrangian density used in the 
calculations. 

The nuclear density has then been folded with the DDM3Y nucleon-nucleon 
interaction to obtain the optical model potential.  
The interaction at distance $r$ for density $\rho$ and projectile energy in centre of 
mass frame $E$, supplemented by a zero range pseudo-potential, is given by
\be v(r,\rho,E)=t^{M3Y}(r,E)g(\rho)\ee
with the M3Y interaction \cite{22,23} in MeV
\be t^{M3Y}=7999\frac{e^{-4r}}{4r}-2134\frac{e^{-2.5r}}{2.5r}-276(1-\frac{E}{200A})\delta(r)\ee
Here $E$ is given in MeV, $r$ in fm and $A$ is the mass number of the projectile.
The density dependent factor is \cite{densdep}
\be g(\rho)=C(1-\beta\rho^{2/3})\ee
 with $C$ and $\beta$ taking values 2.07 and 1.624$fm^2$, respectively,
 obtained from nuclear matter calculation \cite{basu}.
We also use an additional spin-orbit potential $U_{n(p)}^{so}(r)$ 
with energy dependent 
phenomenological potential depths $\lambda_{vso}$ and $\lambda_{wso}$ 
according to Scheerbaum prescription \cite{Scheerbaum}, given by
\begin{equation}
U_{n(p)}^{so}(r)=(\lambda_{vso}+i\lambda_{wso})
\frac{1}{r} \frac{d}{dr} (\frac{2}{3}\rho_{p(n)}+\frac{1}{3}\rho_{n(p)})
\end{equation}
with,
\bea
\lambda_{vso}=130exp(-0.013E)+40\\
\lambda_{wso} =-0.2(E-20) 
\eea

The DDM3Y interaction provides only the real part of the potential. The imaginary part
of the potential is taken to be identical with the real part.
This optical model potential has been used to study the neutron capture reaction cross-sections.

The computer code TALYS1.6 \cite{talys} has been used for cross-section calculations. 
Microscopic level densities, which are important ingredients in statistical model calculations of reaction cross-sections,
are taken from the calculations of Goriely included in the code.
The $\gamma$-strength functions for the dominant E1 $\gamma$ transitions, are taken from Goriely's hybrid model \cite{gor}.
Width fluctuation corrections in compound nuclear decay are also considered. 
These are especially important near threshold energy of new channel openings where channel strengths differ significantly. 
Radial densities are taken from RMF calculations. Pairing energy correction has also been included. At low incident energies,
 {
\em i.e.} below a few MeV, 
mainly binary reactions occur and very often the target and compound nuclei only are involved in the whole reaction chain.
A maximum of 30 discrete levels are taken for both target nuclei and residual nuclei.
Hauser-Feshbach calculations are performed with full $j,l$ coupling. All these options are included in TALYS code. 

\section{Results}
\subsection{Results of RMF calculations}

\begin{figure}[htp]
\vskip -4cm
\includegraphics[scale=0.92]{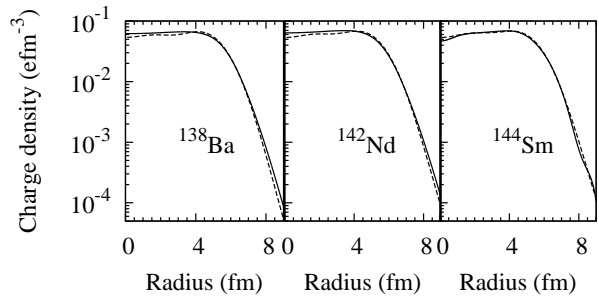}
\caption{Charge density in several $N=82$ nuclei. Solid lines denote the model values 
using the parameters obtained from fitting the experimental data. Dashed lines 
indicate our results.}
\label{density}
\end{figure}

The most important RMF result relevant to the calculation of neutron capture in 
the present formalism is the density profile. Experimental results
on density are available for three
nuclei with $N=82$, {\em viz.} $^{138}$Ba, $^{142}$Nd, and $^{144}$Sm.
In Fig. \ref{density}, we plot the charge density obtained in our calculation
for these three nuclei. The 
experimental densities for $^{138}$Ba and $^{142}$Nd have been generated from 
the parameters for a three parameter Gaussian function fitted to describe
the electron scattering data of Heisenberg \etal\cite{dens1}
For $^{144}$Sm, the Fourier-Bessel coefficients obtained from fitting the 
experimental results of Moinester \etal\cite{dens2} have been used.

\begin{table}[hbt]
\caption{Charge radii of the nuclei studied in the present work. 
Experimental charge radii values are from the compilation of Angeli \cite{rad}.
\label{chrad}}
\begin{tabular}{cccccc}\hline
Nucleus & \multicolumn{2}{c}{$r_c$(fm)}
&Nucleus & \multicolumn{2}{c}{$r_c$(fm)}\\
&Exp. & Pres.&&Exp.&Pres.\\\hline
$^{133}$Cs&4.804&4.801&$^{134}$Cs& 4.803&4.807\\
$^{135}$Cs& 4.807&4.813& 
$^{136}$Cs& 4.806 &4.819\\
$^{137}$Cs& 4.813&4.825\\
$^{130}$Ba&4.829  & 4.797&
$^{132}$Ba&4.831  & 4.808\\
$^{134}$Ba&4.830  & 4.820&
$^{135}$Ba&4.827  & 4.826\\
$^{136}$Ba&4.833  & 4.832&
$^{137}$Ba&4.833  & 4.837\\
$^{138}$Ba&4.838  & 4.843\\
$^{138}$La&4.846  & 4.856&
$^{139}$La&4.855  & 4.862\\
$^{136}$Ce&4.874  & 4.858&
$^{138}$Ce&4.873  & 4.869\\
$^{140}$Ce&4.877  & 4.879&
$^{141}$Ce&       & 4.892\\
$^{142}$Ce&4.906  & 4.905\\
$^{141}$Pr&4.892  & 4.898&
$^{142}$Pr&       & 4.910\\
$^{143}$Pr&       & 4.922\\
$^{142}$Nd&4.912  & 4.915&
$^{143}$Nd&4.923  & 4.927\\
$^{144}$Nd&4.944  & 4.939&
$^{145}$Nd&4.958  & 4.953\\
$^{146}$Nd&4.975  & 4.965&
$^{147}$Nd&4.984  & 4.977\\
$^{147}$Pm&       & 4.981&
$^{148}$Pm&       & 4.993\\
$^{144}$Sm& 4.944 & 4.950&
$^{147}$Sm& 4.984 & 4.985\\
$^{148}$Sm& 5.001 & 4.998&
$^{149}$Sm& 5.011 & 5.010\\
\hline\end{tabular}
\end{table}

Charge radius values, as a first 
moment
of the charge distribution, can be compared with calculated results to
check the success of theoretical approach. In Table \ref{chrad}, we compare the
calculated r.m.s. charge radius  values with experimental data. From 
the comparison between experimental and calculated charge density and radius 
values, one can infer that the present RMF calculation can describe the 
nuclear density  near $N=82$ shell  closure very well. We now employ the
theoretical density values to derive the optical model  potential
and extract cross-sections for neutron capture.

\subsection{Cross-sections at various neutron energies}

TALYS-1.6 is meant for analysis of data above the resolved resonance range,
which is approximately 1 keV. We have compared the neutron capture 
cross-sections with those experimental measurements which do not show resolved 
resonances
at low energy.

\begin{figure}[htp]
\includegraphics[scale=0.62]{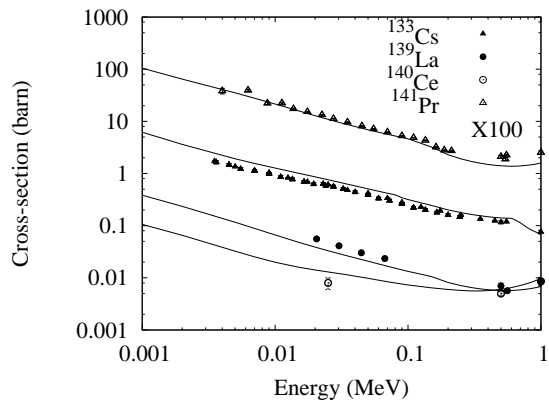}
\caption{Comparison of results of the present calculation with experimental measurements
for $^{133}$Cs, $^{139}$La, $^{140}$Ce, and $^{141}$Pr. The solid lines indicate the theoretical 
results. For convenience, cross-section values for $^{141}$Pr have been multiplied by a factor of 100.
\label{Cs}}
\end{figure}

\begin{figure}[htp]
\includegraphics[scale=0.62]{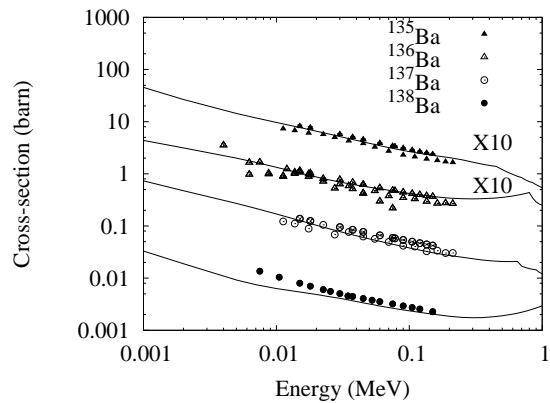}
\caption{Comparison of results of the present calculation with experimental measurements
for $^{135-138}$Ba. The solid lines indicate the theoretical results. For convenience,  
cross-section values for $^{135,136}$Ba have been multiplied by a factor of 10. 
\label{Ba}}
\end{figure}

\begin{figure}[htp]
\includegraphics[scale=0.62]{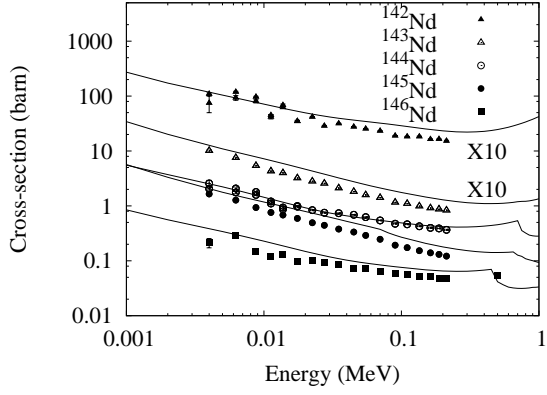}
\caption{Comparison of results of the present calculation with experimental measurements
for $^{142-146}$Nd. The solid lines indicate the theoretical results. For 
convenience, cross-section values for $^{143,144}$Nd have been multiplied by a factor of 10.
\label{Nd}}
\end{figure}

\begin{figure}[htp]
\includegraphics[scale=0.62]{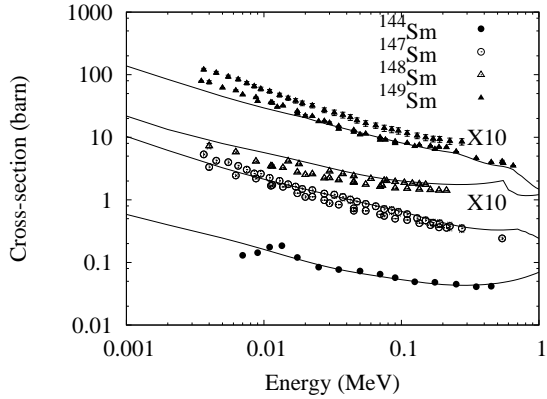}
\caption{Comparison of results of the present calculation with experimental measurements
for $^{144,147-149}$Sm. The solid lines indicate the theoretical results. For convenience,
cross-section
values for $^{148,149}$Sm have been multiplied by a factor of 10.
\label{Sm}}
\end{figure}

Theoretical neutron capture cross-section results for different neutron energy 
values
have been compared with experiments in Figs. \ref{Cs} - \ref{Sm}.
In general, we have considered the more recent results. Our main interest lies in the
nuclei which are important for astrophysical s- and p-processes. We 
 generally present only those results where reasonable amount of experimental data exist for 
comparison. The exceptions are $^{139}$La and $^{140}$Ce because these three nuclei correspond to
the  $N=82$ shell closure.

For convenience, 
we present the results for the elements, where only one isotope have been 
studied, in Fig. \ref{Cs}. These include $^{133}$Cs, $^{139}$La, $^{140}$Ce, and $^{141}$Pr.  
Experimental cross-section values for $^{133}$Cs($n,\ga)$ reaction are from Refs.
 \cite{Cs-1,Cs-2,Cs-3}. The cross-section values are averaged over neutron energy range as the resonances 
are unresolved. For example, Yamamuro \etal \cite{Cs-2} used the neutron beam from a tantalum 
photo-neutron source. They then average the cross-section over appropriate energy
interval. 
For $^{139}$La, the data are from Refs. \cite{Cs-3,La-1}.  Similarly, here also the neutron 
energy has a 5\% error and we get an average value for various
energies.
Harnood \etal measured the neutron capture cross-section of $^{140}$Ce and $^{141}$Pr using the neutron TOF technique \cite{Ce}.
Voss \etal \cite{Pr141-1} also 
measured the cross-section for $^{141}$Pr in the range 3 to 225 keV. They 
also calculated the MACS values from their results.
Voignier \etal also measured the capture cross-section in the energy range 0.5 to 3 MeV \cite{Pr141-2}.

In Fig. \ref{Ba}, we show the average neutron capture cross-sections in 
$^{135-138}$Ba nuclei and compare with experimental values between 1 keV and 1 MeV.
Experimental cross-section values are from Refs. \cite{Voss,Voss1,Beer, Beer1}.
Voss \etal  studied the neutron capture cross-section
for $^{134-137}$Ba nuclei in the energy range from
3 to 225 keV using gold as a standard \cite{Voss,Voss1}.  Neutron capture
by the neutron closed shell nucleus $^{138}$Ba has been studied
in Refs. \cite{Beer,Beer1}.

Results for  neutron capture cross-sections of $^{142-146}$Nd are shown in 
Fig. \ref{Nd}.
Wisshak \etal \cite{Wisshak1,Wisshak2} studied the resonances above 3 keV in 
$^{142-146,148}$Nd. The data were then compressed in 
coarse bins to get the average behaviour. In another work, Veerapaspong 
\etal studied the neutron capture cross-sections for $^{143.145,146}$Nd 
 \cite{Nd-2}.
Their data were put in a large energy bin of 10 keV wide. Their results agree
with Refs. \cite{Wisshak1,Wisshak2} and our results though we have not shown 
the data in the figure.

 In Fig. \ref{Sm}, we plot the experimental and calculated values for 
$^{144,147,148}$Sm.
 For $^{144}$Sm, experimental values are from Macklin \etal
\cite{Sm-1} where the authors have made measurements from 0.5 eV to 500 keV and have obtained the resonance parameters up to 100 keV. 
Wisshak \etal studied the neutron capture cross-section of $^{147-150,152}$Sm
in the energy range 3 to 225 keV using gold as standard \cite{Sm-2}. Mizumoto \cite{Sm-3} measured
the neutron capture cross-sections of $^{147,149}$Sm in the energy range 3 - 300
keV using TOF technique. Diamet \etal used a similar technique to study
$^{147-150,152,154}$Sm in the energy range 10 to 90 keV \cite{Sm-4}.

From Figs. \ref{Cs} - \ref{Sm}, one can see that the theoretical calculation
reproduces the experimental values in most of the cases.  In the next subsection, we employ our 
method to calculate the MACS for some astrophysically important nuclei.

\subsection{Maxwellian averaged cross-sections}

\begin{figure}[htb]
\includegraphics[scale=0.64]{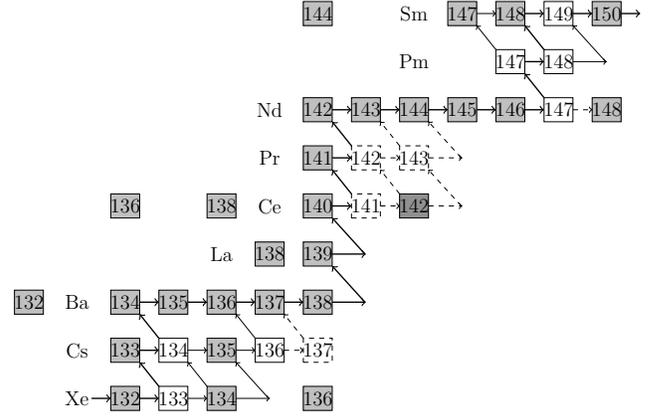}
\caption{\label{s-path}The s-process path near the shell closure at $N=82$. Also
shown are some of the p-nuclei. 
See text for more details.}
\end{figure}

\begin{table*}[htbc]
\caption{Maxwellian averaged cross-sections at $kT=30$ keV for nuclei near the 
$N=82$ shell closure. Experimental values are from Ref. \cite{kadonis}. See text for other 
experimental values. The nuclei with $N=82$ are in bold font.
\label{macs}}
\begin{tabular}{crrrcrrr}\hline
 &\multicolumn{3}{c}{MACS(mb)}&
 &\multicolumn{3}{c}{MACS(mb)}\\\cline{2-4}\cline{6-8}
Nucleus&Pres. & Exp. & MOST &Nucleus&Present & Exp. & MOST\\\hline
$_{55}^{133}$Cs&     685  &    509$\pm$21    &  469&
$_{55}^{134}$Cs&     786     &     & 805\\
$_{55}^{135}$Cs&     147 &     160$\pm$10   &   148& 
$_{55}^{136}$Cs&  90.4     &     & \\
$_{55}^{137}${\bf Cs}&    16.1   & &\\
$_{56}^{130}$Ba&    625.2  &     746$\pm 34$    &  490&
$_{56}^{132}$Ba&     393  &     397$\pm 16$    &  227\\
$_{56}^{134}$Ba&     158     &   176.0$\pm$ 5.6    &117& 
$_{56}^{135}$Ba&     528      &  455$\pm$ 15    &  259\\
$_{56}^{136}$Ba&     74.6     &  61.2$\pm$2.0   &   49.4& 
$_{56}^{137}$Ba&     81.8    &   76.3$\pm2$.4   & 95.4\\
$_{56}^{138}${\bf Ba}&     4.14   &     4.00$\pm$0.20  &  2.79\\
$_{57}^{138}$La&     417    &        & 337& 
$_{57}^{139}${\bf La}&     31.0    &   32.4$\pm$3.1   & 45.9 \\ 
$_{58}^{136}$Ce&     547     &   328$\pm 21$    & 206& 
$_{58}^{138}$Ce&     137     &   179$\pm5$    &  60.5\\
$_{58}^{140}${\bf Ce}&     12.7   &    11.0$\pm$0.4    & 6.71& 
$_{58}^{141}$Ce&    198.8    &        & 58.4 \\
$_{58}^{142}$Ce&     33.5    &   28$\pm 1$     &  16.7\\
$_{59}^{141}${\bf Pr}&     101    &    111.4$\pm$1.4  & 130& 
$_{59}^{142}$Pr&     233    &      & 261\\
$_{59}^{143}$Pr&     170    &       &  57.7\\
$_{60}^{142}${\bf Nd}&     54.5   &    35.0$\pm$0.7    &  22.9& 
$_{60}^{143}$Nd&     362.4   &   245$\pm$3     &  105\\
$_{60}^{144}$Nd&     82.3    &   81.3$\pm$1.5   &  37.1& 
$_{60}^{145}$Nd&    617.9     & 425$\pm $5      & 207  \\
$_{60}^{146}$Nd&    128.7 &  91.2$\pm$1.0    & 56.8  & 
$_{60}^{147}$Nd&    1434.4 &      & 663 \\
$_{61}^{147}$Pm&   1210.4  &  709$\pm$100    & 452  & 
$_{61}^{148}$Pm&  1529.7  &      &  \\
$_{62}^{144}${\bf Sm}&     91.0   &    $92\pm$ 6      &  38.6& 
$_{62}^{147}$Sm&    1229    &    973$\pm$10    &  584\\
$_{62}^{148}$Sm&     340    &    241$\pm$2     &  130& 
$_{62}^{149}$Sm&   1622     &    1820$\pm$17     & 1274 \\
\hline
\end{tabular}
\end{table*}
Apart from the nuclei with $N=82$ in the s-process path, {\em i.e.} $^{138}$Ba,
$^{139}$La, $^{140}$Ce, $^{141}$Pr, and $^{142}$Nd, which act as bottlenecks due
to low cross-sections, 
neutron capture reactions in 
some other nuclei in the neighbourhood are also important for nucleosynthesis.
In Fig. \ref{s-path}, we show the s-process path in the neighbourhood of  
the shell closure at $N=82$. 
The shaded rectangles indicate stable and 
extremely long-lived nuclei. 
The weak branch points are indicated by rectangles with 
dashed lines. Similarly, weak s-process paths are indicated by dashed lines.
One can see that there are strong branch points in the s-process path at $^{134}$Cs, 
$^{147}$Nd, and $^{147,148}$Pm. Besides, 
there are weaker branch points at $^{136}$Cs, $^{141}$Ce, and $^{147,149}$Pm. As already
pointed out, the nuclei $^{134,136}$Ba, $^{142}$Nd, and $^{148,150}$Sm
are s-only.  There are several p-nuclei
such as $^{130}$Ba (not shown in Fig. \ref{s-path}) $^{132}$Ba, $^{138}$La, 
$^{136,138}$Ce, and $^{144}$Sm, the last corresponding to the shell closure.

In Table \ref{macs}, we present the theoretically calculated  MACS values at 
$kT=30$ keV for a number of selected 
isotopes and compare with experimental measurements whenever available. 
Experimental values are from the database\cite{kadonis},
which is an updated version of the compilation of Bao \etal\cite{bao}.
Theoretical values from MOST calculation \cite{MOST,MOST1}, which are listed on the KADONIS
online database, are also presented. MOST is a Hauser-Feshbach code which derives
all nuclear inputs from global microscopic models.
As one can see, agreement of our results  with experiment is better than the MOST results 
in almost all cases.
In the next part of the discussion, we comment only on the more significant 
results.

The Cs isotopes that are involved in the s-process are $^{133,134,135}$Cs.
The nucleus $^{134}$Cs is unstable with a half life of 2.06 years  and no 
neutron capture data is yet 
available. 
However, this is an important for the abundance of $^{134,136}$Ba in view of the 
strong branching at $^{134}$Cs. The compilation by Bao \etal \cite{bao} recommended
a MACS value of 664$\pm 174$ mb. Our calculated value for $^{135}$Cs is
close to the experimental measurements. However, the neutron deficient
$^{133}$Cs rate is comparatively poorly reproduced. 

Dillmann \etal \cite{p-proc} have measured the MACS for a number of nuclei
relevant to p-process. The measured values  at $kT=25$ keV for 
the nuclei $^{130,132}$Ba are 736$\pm 29$ mb and 392.6$\pm 14.8$ mb, 
respectively. 
Our calculated value for $^{130,132}$Ba at 25 keV are 675.3 and 424.1 mb,
respectively.
The neutron capture cross-sections
of the s-only nuclei  $^{134,136}$Ba are very important for constraining the s-process. The  
cross-section for $^{136}$Ba is also known to be an important ingredient in determining the mean
neutron exposure in the main s-process component.
As one can see, here also our results are reasonably close to experimental measurements
except in the case of $^{130}$Ba.

\begin{table*}
\caption{Maxwellian averaged neutron capture cross-sections of
$^{138}$Ba, $^{139}$La, $^{140}$Ce, $^{141}$Pr, and $^{142}$Nd.
The experimental values presented are from Ref. \cite{kadonis}. See text for
more details about the experiments and the
available latest measurements not included in Ref. \cite{kadonis}.
\label{la139} }
\begin{tabular}{crrrrrrrrrr}\hline
$kT$(MeV) & \multicolumn{10}{c}{MACS (mb)}\\
&\multicolumn{2}{c}{$^{138}$Ba}&\multicolumn{2}{c}{$^{139}$La}
&\multicolumn{2}{c}{$^{140}$Ce}&\multicolumn{2}{c}{$^{141}$Pr}
&\multicolumn{2}{c}{$^{142}$Nd}\\
& Pres. & Expt.& Pres. & Expt.& Pres. & Expt.
& Pres. & Expt.& Pres. & Expt.\\\hline
0.005& 10.3 &13.4  &118  & 111.2&
34.0 &23&353&412&146.5&98.6\\
0.010& 6.96& 7.85  &68.9 &   63.2&21.9&19.5&215&247&95.8&65.1\\
0.015& 5.68& 5.93  &50.8 &   48.1&17.6&16&161&182&76.2&51.3\\
0.020& 4.93& 4.95  &41.2 &   40.3&15.2&13.5&132&148&65.8&43.4\\
0.025& 4.46& 4.38  &35.2 &    35.7
&13.8&12&114&126&58.8&38.4\\
0.030& 4.14& 4.00  &31.0 &    32.4
&12.7&11.0&101&111&54.5&35.0\\
0.040& 3.61& 3.49  &25.4 &    27.7&11.2&9.5&82.8&91.5&47.9&30.7\\
0.050& 3.28& 3.14  &21.8 &    23.6&10.2&8.7&70.8&78.3&43.8&27.7\\
0.060& 3.08& 2.89  &19.2 &   22.0&9.5&8.1&62.0&69.0&41.0&25.5\\
0.080& 2.76& 2.52  &15.8 &   18.4&8.6&7.2&50.1&56.2&37.4&22.9\\
0.100& 2.58& 2.23  &13.7 &   15.2&8.0&6.6&42.4&47.6&35.5&21.0\\
\hline
\end{tabular}
\end{table*}

K\"{a}ppeler \etal measured the cross-sections for 
stable Ce isotopes \cite{Ce140}. They then constructed an optical model potential
for this region and calculated the cross-sections for $^{141}$La and $^{142,143}$Pr
in the Hauser-Feshbach formalism. Their calculated values for these two nuclei
at 30 keV are 91 mb, 297 mb, and 205 mb, respectively. Although their results for
$^{140,142}$Ce and $^{141}$Pr are very close to our calculations, the value for $^{141}$Ce
is smaller by more than a factor of two while that for $^{142,143}$Pr is larger by more than
25\% and 20\%, respectively.
The results for La and Pr isotopes are also close to experimental measurements.
The nucleus $^{138}$La is produced in the p-process. However, results for more
neutron deficient Ce isotopes do not agree well with experiments.

As one goes to heavier isotopes, agreement becomes poorer except in a few cases
such as $^{144}$Nd and $^{144}$Sm, though they are still better than the MOST calculation. 
In general, the poor agreements may be due to the fact that away from the closed shell,
deformation effects come into the picture. However, our calculation is unable 
to explain 
the recent results for $^{142}$Nd, a spherical nucleus with $N=82$, though 
some of the
older
measurements for $^{142}$Nd are closer to our calculation. Results for all the other 
nuclei with $N=82$ are explained with a good accuracy. 

As already pointed out, modern measurements have emphasized the importance 
of the MACS  values at various thermal energies. Hence a number of works, apart 
from the already mentioned, have measured the MACS values at different 
temperatures. We present the MACS values at different temperatures
for $N=82$ isotopes in Table \ref{la139} and draw attention to some important 
results. 
Heil
\etal \cite{Ba138-1} have used neutron activation studies to measure the
MACS value at $5.1$ keV as $13.0\pm 0.5 $ mb in $^{138}$Ba. 
In the present work, MACS for $^{138}$Ba is calculated to be 10.2 mb at $kT=5.1$ keV. 

Natural lanthanum is nearly mono-isotopic.
It is an important element as it can be easily detected in solar spectroscopy.
It is produced in both s- and r-processes 
and is particularly suitable for monitoring s-process abundances from Ba
to Pb. 
This has led to the study of $^{139}$La through neutron TOF 
spectroscopy as well as activation measurement. 
In Table \ref{la139}, we present the MACS values at different temperatures. 
O'Brien \etal have measured the Maxwellian averaged cross-section (MACS) at
$kT= 30$ keV as 31.6$\pm0.6$ mb \cite{La139-1}. 
Our calculated value of $31.0$ mb agrees with the measurements. Activation
technique has also been used to measure MACS at $kT=5$ keV. The measured
value 113.7$\pm$4.0 \cite{La139-2} is in excellent agreement with our calculation.
Terlizzi \etal have measured the resonance parameters in the energy range
0.6 eV to 9 eV and have recalculated the MACS values in the light of their measurements \cite{La139-3}.
Their values also lie close to our calculated results. 
For example, their measurement yields a value of $106.9\pm5.3$ mb at
$kT=5$ keV after  
normalization to the value for 25 keV from Ref. \cite{La139-1}.

K\"{a}ppeler \etal \cite{Ce140}  also used the $^7$Li($p,n$)$^7$Be reaction
to study the thermal neutron capture by Ce isotopes at 25 keV.
They obtained a MACS value of 12.0$\pm 0.4$ mb. This was  extrapolated
to other thermal energy values. As already mentioned, 
Voss \etal\cite{Pr141-1} have obtained the MACS values as a function of temperature 
for $^{141}$Pr.

As already mentioned, Wisshak \etal\cite{Wisshak1} studied the neutron 
capture
cross sections of Nd isotopes. They have also calculated the MACS values at 
different energy from their data. Guber \etal \cite{Guber} also 
measured the MACS values between $kT$ = 5 - 50 keV. Both the above references 
have commented on the importance of the 
new measurements in s-process.

\section{Summary}

To summarize, astrophysically important neutron capture reactions near the $N=82$ 
shell closure have been studied using a microscopic approach. 
Densities of relevant nuclei near have been calculated in the
RMF approach. The calculated charge densities and radii agree with experimental 
measurements, whenever available. The calculated density has been folded with 
DDM3Y nucleon-nucleon interaction to obtain the optical model potential for
neutron reactions. Cross sections for $(n,\ga)$ reactions have been calculated and compared 
with measurements. Finally Maxwellian average cross-section values, important for
s- and p-processes, have been calculated.

\section*{Acknowledgment}

The authors acknowledge the financial support provided by
University Grants Commission (India), Department of Science
and Technology, Alexander Von Humboldt Foundation, and
the University of Calcutta.

\end{document}